\documentstyle[aps,epsfig]{revtex}
\begin{document}

\draft
\narrowtext
\twocolumn
\wideabs{
\title{Quantum anti-centrifugal force}
\author{M. A. Cirone$^{a}$, K. Rz\c{a}\.{z}ewski$^{b}$,
W. P. Schleich$^{a}$, F. Straub$^{a}$, J. A. Wheeler$^{c}$}
\address{$^{a}$ Abteilung f\"{u}r Quantenphysik,
Universit\"{a}t Ulm, D-89069 Ulm, Germany \\
$^{b}$ Center for Theoretical Physics, Polish Academy
of Sciences, and College of Science, \\
Al. Lotnik\'{o}w 32/46, 02-668 Warsaw, Poland \\
$^{c}$ Princeton University, Department of Physics and Astronomy,
Princeton, NJ 08544, USA}

\date{\today}
\maketitle

\begin{abstract}
In a two-dimensional world a free quantum particle of vanishing
angular momentum experiences an attractive force. This force originates
from a modification of the classical centrifugal force due to
the wave nature of the particle. For
positive energies the quantum anti-centrifugal force manifests itself
in a bunching of the nodes of the energy wave functions towards
the origin. For negative energies this force is sufficient to
create a bound state in a two-dimensional delta function potential.
In a counter-intuitive way the attractive force pushes the particle
away from the location of the delta function potential. As a
consequence, the particle is localized in a band-shaped domain
around the origin.
\end{abstract}

}

\vspace{2.0cm}

\twocolumn
\narrowtext

\section{Introduction}

`If this is the best of all possible worlds, what
are the others like?' exclaims Candide \cite{cand} in
Voltaire's philosophical novel when he sees the
devastating results of the earthquake in Lisbon.
Almost 150 years after Voltaire, Einstein pondered
the question
`How much freedom had God when he created the world?'.
In the same spirit P. Ehrenfest \cite{ehr,tip} raised the problem
`Why is the space we live in three-dimensional?'.
Since then many phenomena where dimensionality of
space plays a crucial role have been discovered.
They manifest themselves in quantum dots and wires in solid state
physics, phase transitions in statistical physics or
in the Kaluza-Klein or string theories of particle physics.
In the present paper
we point out a wave effect that is unique to two
space dimensions and that can, in principle, be
observed in the recent two dimensional trapping experiments
using wires \cite{den}: A point particle subjected to a
potential which is solely confined to the coordinate origin
binds locally in one and three dimensions but
in two dimensions binds in a domain like
a hollow pipe. The deeper reason for this surprising effect
lies in the quantum anti--centrifugal potential: In two
dimensions the centrifugal potential corresponding
to vanishing angular momentum is attractive
rather than repulsive.

In the present paper we focus on the manifestations of
the quantum anti--centrifugal potential in the
energy eigenstates of a free particle in two
dimensions. The problem of time--dependent phenomena
originating from this potential will
be addressed in future publications.

The paper is organized as follows:
In Sec. \ref{anun} we observe that a localized wave
function satisfies the time independent Schr\"odinger
equation of a free particle. The reason for the
localization stands out most clearly in the
Schr\"odinger equation for the radial wave function, discussed
in Sec. \ref{anat}. Indeed, for vanishing angular momentum
an attractive potential arises from the
wave nature of the particle and determines the decay of
the radial wave function. In Sec. \ref{node} we identify the
origin of
the corresponding attractive force as interference
of waves. Moreover, we show that the attraction or repulsion of
the potentials corresponding to vanishing or one unit
of angular momentum manifests itself in the bunching
or anti--bunching of the nodes of the radial
wave function. This phenomenon of attraction is unique to
two dimensions. In Sec. \ref{boun} we address the question
of a bound state of a `free' particle. Indeed, the attraction
due to the quantum anti--centrifugal force is not enough to
create a bound state. An additional weakly binding
potential, such as a delta function potential, is
necessary. We conclude in Sec. \ref{conc} by presenting
some ideas for experimental realizations of these
considerations.

\section{An unusual bound state}
\label{anun}

Our analysis rests on the observation that
the function \cite{berry,col}

\begin{equation}
\Phi^{(2)}(x,y) \equiv \frac{1}{\sqrt{\pi}}\: k\: K_{0}
\left( k\: \sqrt{x^2+y^2}\right)
\label{eq1}
\end{equation}
defined in terms of the zero-th modified Bessel
function $K_{0}$ and the wave number $k$
satisfies the Helmholtz equation 

\begin{equation}
\left[ \Delta^{(2)}-k^2 \right] \Phi^{(2)}(x,y)=0,
\label{eq2}
\end{equation}
everywhere except at $x=y=0$. Here,
$\Delta^{(2)}$ denotes the Laplacian in
two dimensions.

When we recall the dispersion relation

\begin{equation}
E=-\mid E \mid = -\frac{(\hbar k)^2}{2M},
\label{disp}
\end{equation}
of a free particle with mass $M$ and negative energy
$E$, the Helmholtz equation is equivalent to the
corresponding time independent Schr\"{o}dinger equation.

\begin{figure}
\begin{center}
\epsfig{width=3.5in,file=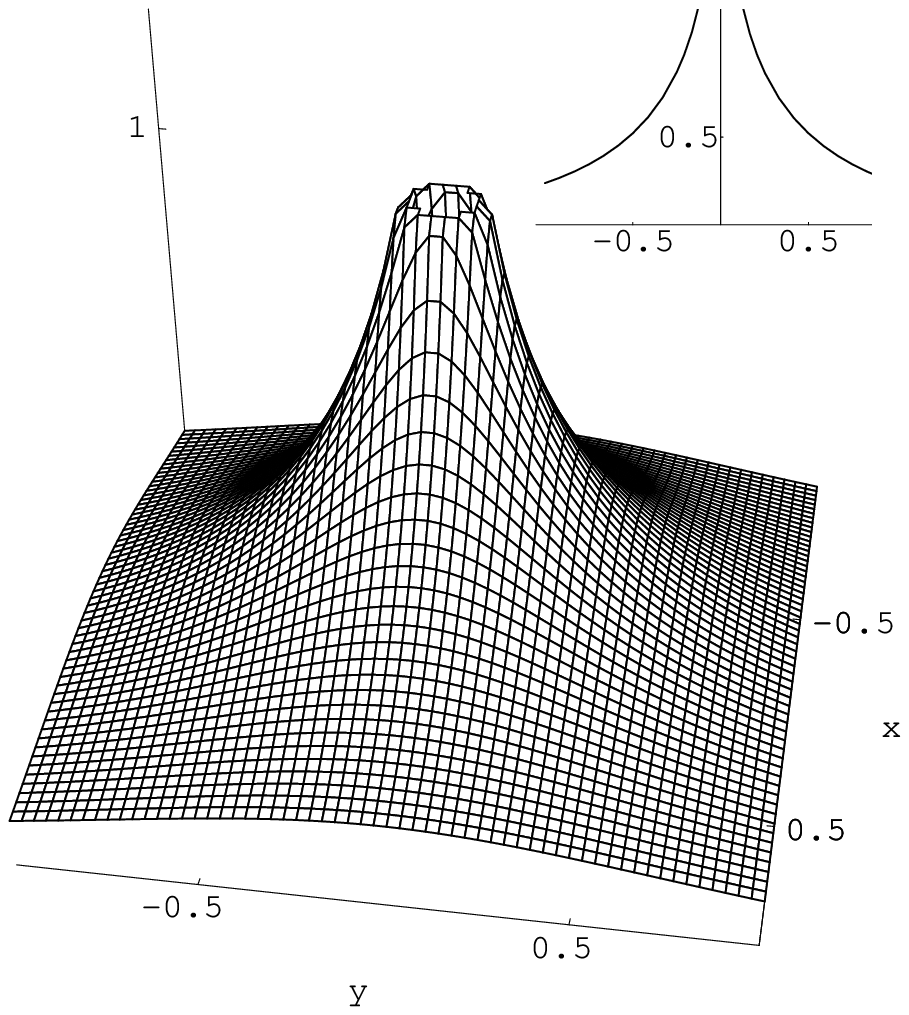,angle=0}
\caption{The wave function $\Phi^{(2)}(r)$ represented
in two--dim\-ensional space is logarithmically
divergent at the origin but decays exponentially for
positions away from the origin. In the insert
we show a cut along the $x$ axis which brings
out the logarithmic divergence of $\Phi^{(2)}(r)$
at the origin.}
\end{center}
\label{wave}
\end{figure}

The wave function $\Phi^{(2)}$ shown in Fig.\ref{wave}
enjoys some rather
unusual properties: Due to the modified Bessel function $K_{0}$,
it diverges \cite{abrste} logarithmically at the origin
whereas at large distances it decreases exponentially.
Despite this divergence, the wave function
is still square integrable,

\begin{eqnarray*}
\int_{-\infty}^{\infty} \!\!\!\!\!\! dx
\int_{-\infty}^{\infty}\!\!\!\!\!\! dy
\mid \Phi^{(2)}(x,y)\mid^2 & = & \int_{0}^{\infty} \!\!\!\!\!\!dr \!\!
\int_{0}^{2\pi}\!\!\!\!\!\! r\:d\phi\: \frac{k^2}{\pi}K_{0}^2(kr)
\nonumber \\
& = & 2\int_{0}^{\infty}\!\!\!\!\!\!\! d\xi \: \xi K_{0}^2(\xi) = 1.
\end{eqnarray*}

Indeed, the area element
$dx\:dy=r\:dr\:d\phi$ brings in an additional
power of $r \equiv (x^2+y^2)^{1/2}$
and regularizes the
logarithmic divergence at the origin.

\begin{figure}
\begin{center}
\epsfig{width=3.5in,file=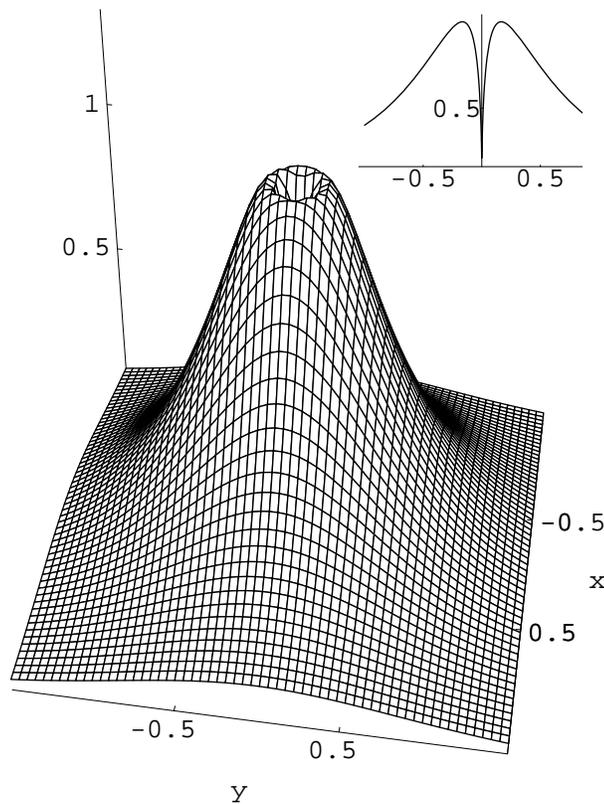,angle=0}
\caption{The radial probability $W^{(2)}(r)$ vanishes at the origin
and decays for large distances, with a maximum close to the
origin. In the insert we show a cut along the $x$ axis
which brings out the cusp of $W^{(2)}(r)$ at the origin.}
\end{center}
\label{probab}
\end{figure}

For the same reason, the probability

\begin{displaymath}
W^{(2)}(r)\: dr \equiv 2k^2K_{0}^2(kr)\: r\: dr
\end{displaymath}
to find the particle between $r$ and $r+dr$
vanishes at the origin, as shown in Fig.\ref{probab}.
Moreover, since the
modified Bessel function
$K_{0}$ decays for large distances the radial
probability displays a maximum close to the origin.

\section{Quantum anti--centrifugal potential}
\label{anat}

What is the deeper reason for this localization \cite{rau,walt}
of a free particle? No classical potential
prevents the particle from diffusing away.
One part of the answer to this apparently paradoxical
situation -- a bound state of a free particle --
lies in the Schr\"{o}dinger equation

\begin{displaymath}
\left\{ \frac{d^2}{dr^2}+\frac{2M}{\hbar^2}
\left[ E-V_{m}^{(2)}(r)\right] \right\} u^{(2)}_{m}(r)=0
\end{displaymath}
for the radial wave function

\begin{equation}
u^{(2)}_{m}(r) \equiv \sqrt{r}e^{-im\phi}\Phi^{(2)}
(r\cos \phi, r\sin \phi).
\label{rad}
\end{equation}
Here we have introduced the effective potential

\begin{displaymath}
V_{m}^{(2)}(r)\equiv \frac{\hbar^2}{2M}
\frac{m^2-1/4}{r^2}
\end{displaymath}
in two dimensions.
The radial wave equation Eq.(\ref{rad})
follows from the Helmholtz equation
Eq.(\ref{eq2}) with the help of the dispersion
relation Eq.(\ref{disp}).

\begin{figure}
\begin{center}
\epsfig{width=4.5in,angle=0,file=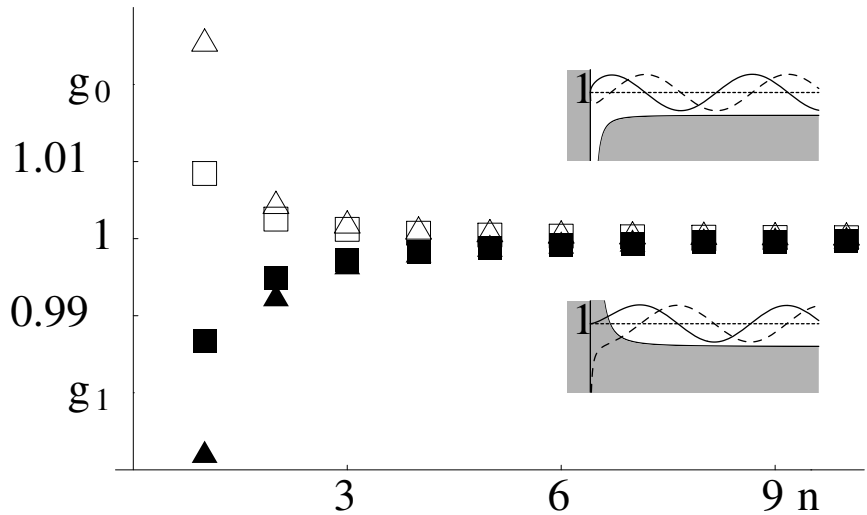}
\caption{Node bunching and anti-bunching of energy eigenfunctions
of a free particle in a two-dimensional space.
The centrifugal potential corresponding
to a non-vanishing angular momentum is repulsive (bottom inset) and
the two linearly independent eigenfunctions are determined
by the Bessel function $J_{1}$ (solid line) and the Neumann function $Y_{1}$
(dotted line). In contrast, the
potential corresponding to a vanishing angular momentum is
attractive (top inset) and the two eigenfunctions are proportional to
$J_{0}$ (solid line) and $Y_{0}$ (dotted line).
The repulsive and attractive potentials give rise to an anti-bunching and
bunching of the nodes of the energy eigenfunction, respectively.
As a
measure $g_{m}(n) \equiv \pi / \Delta_{m}(n)$ of bunching or anti-bunching we
use the inverse of the difference $ \Delta_{m}(n)$
of neighbouring zeros of the $m$-th Bessel function $J_{m}$
or Neumann function $Y_{m}$ in units
of the free space separation $\pi$.
Filled squares or triangles represent $g_{1}(n)$ for $J_{1}$ or $Y_{1}$
in the repulsive centrifugal potential. Open squares or triangles
represent $g_{0}(n)$ for $J_{0}$ or $Y_{0}$ in the attractive potential.
The zeros of $Y_{0}$ and $Y_{1}$ lie closer to
the origin than those of $J_{0}$ and $J_{1}$.
Consequently, the bunching or anti--bunching effect is more evident in the
Neumann function than in the Bessel function. The physics of the
non-relativistic free particle does not contain an intrinsic unit of length.
When we define a dimensionless  length $\rho \equiv k r$ where $k$ is the wave
number, the dimensionless energy eigenvalue is unity.}
\end{center}
\label{bun}
\end{figure}

The first term in $V_{m}^{(2)}$,
proportional to $m^2$, is the potential
which describes the familiar centrifugal force. Less
familiar is the negative correction term $-1/4$ which
comes from the reduction of space from three to two
dimensions. It gives rise to a centripetal force
which from this point on we shall call a quantum
anti-centrifugal force to emphasize that its binding
power arises from quantum mechanics.
Indeed, for
particles with nonvanishing angular momentum
($m\neq 0$) the potential is repulsive, as shown
in the bottom inset of Fig.\ref{bun}. However, the
repulsiveness associated with the classical centrifugal
force, that is the $m^2$ term, is softened by the
correction term $-1/4$.

The effect of this contribution
stands out most clearly for
a particle with zero angular momentum, that is $m=0$.
Here the effective potential shown in
the top inset of Fig.\ref{bun} becomes attractive. Hence,
this quantum anti-centrifugal potential

\begin{displaymath}
V_{Q}(r) \equiv V_{0}^{(2)}(r)
= -\frac{\hbar^2}{2M}\frac{1}{4r^2}.
\end{displaymath}
is the reason for the decay of the
wave function Eq.(\ref{eq1}) at large distances.

We have chosen this name for the potential
to bring out in the most striking way the counterintuitive
nature of this attraction. However, we emphasize
that, despite the name, the attraction is not related
to the angular but to the radial motion.

To illustrate this statement
we compare the effective potential
$V^{(2)}_{m}$ in
two dimensions to the effective potential

\begin{equation}
V_{l}^{(3)}(r)\equiv \frac{\hbar^2}{2M}
\frac{l(l+1)}{r^2}
\end{equation}
in three dimensions. Here, $l$ denotes the quantum number
of angular momentum.

Both potentials seem to be quantum translations of
the classical centrifugal potential

\begin{equation}
V_{\rm cl}(r) \equiv \frac{\vec{L}^2}{2Mr^2}
\end{equation}
where $\vec{L}$ is the angular momentum vector.
Indeed, in three dimensions the
`quantum square' of angular momentum
reads $\vec{L}^2=\hbar^2l(l+1)$. In
two dimensions it seems to take the less familiar form
$\vec{L}^2=\hbar^2 (m^2-1/4)=\hbar^2(m-1/2)(m+1/2)$.

However, this picture is misleading. Whereas the
quantum square $l(l+1)$ is solely a consequence
of the angular momentum algebra, the correction
term $-1/4$ in two dimensions does not result from
the angular motion, but from the
radial motion. It can be traced back to the
radial derivatives in the Laplacian

\begin{equation}
\Delta^{(2)}=\frac{\partial^2}{\partial r^2}
+\frac{1}{r}\frac{\partial}{\partial r}
+\frac{1}{r^2}\frac{\partial^2}{\partial \phi^2}
\end{equation}
expressed in polar coordinates.

This feature suggests that the quantum
anti--centrifugal force is a metric force.
It originates from the use of curvilinear 
coordinates, that is, the description of the wave in
cylindrical coordinates.

\section{Node bunching and anti--bunching}
\label{node}

How can we gain some insight into the physical origin
of the quantum anti--centrifugal potential $V_{Q}$?
One strategy is to first
consider the familiar case of a free particle of
positive energy and compare and contrast the wave
functions
of an attractive potential ($m=0$) and a repulsive
potential ($m>0$). Then we extend these considerations to
negative energies and emphasize the uniqueness of
two dimensions.

\subsection{Positive energy}

For $E>0$ the two linear
independent solutions
of the two-dimensional Helmholtz equation are the
ordinary Bessel functions $J_{m}$ and the Neumann
functions $Y_{m}$.
For the attractive potential $V_{Q}$
the independent solutions are proportional to $J_{0}$ or $Y_{0}$,
whereas for the repulsive potential $V_{1}^{(2)}$,
corresponding to
$m=1$, we find $J_{1}$ and $Y_{1}$. The different nature of the
potentials -- attractive versus repulsive --
manifests itself in the wave functions through the
distribution of nodes determined by the zeros $j_{m,n}$ or
$y_{m,n}$
of the Bessel function $J_{m}$ or the Neumann function $Y_{m}$.
A measure for the distribution of nodes is the normalized density

\begin{equation}
g_{m}(n) \equiv \frac{\pi}{\Delta_{m}(n)},
\end{equation}
of the zeros of the Bessel functions. Here,

\begin{equation}
\Delta_{m}(n)\equiv j_{m,n+1}-j_{m,n} 
\end{equation}
denotes the separation of neighbouring zeros of $J_{m}$ and

\begin{equation}
\Delta_{m}(n)\equiv y_{m,n+1}-y_{m,n} 
\end{equation}
denotes the same quantity for $Y_{m}$. We have normalized the
separation to the free space separation $\pi$ of the zeros.

In $J_{0}$ and $Y_{0}$
the separation $\Delta_{0}(n)$
between neighbouring zeros decreases for decreasing $n$,
in agreement with the intuitive picture that the particle
accelerates towards the origin. In Fig.\ref{bun}
we represent by open squares and triangles the normalized
density $g_{0}(n)$ of the zeros
of $J_{0}$ and $Y_{0}$, respectively, clearly
demonstrating node bunching.

In the language of cold atoms
the energy wave function $u^{(2)}_{0}$ has a negative
scattering length, indicating an attractive potential.
In the case of cold atoms
the origin of this attraction is a physical interaction.
In contrast, the attractive quantum anti--centrifugal
potential is not due to a classical
interaction but arises from the wave equation.

In contrast,
in $J_{1}$ and $Y_{1}$ the separation $\Delta_{1}(n)$
of neighbouring zeros increases as $n$ decreases,
corresponding to a deceleration
of the particle running up the potential well.
Again, in the language of cold atoms this case corresponds
to a positive scattering length.
The filled squares and triangles of Fig.\ref{bun},
corresponding to the normalized density $g_{1}(n)$
of zeros of $J_{1}$ and $Y_{1}$, respectively,
reflect the phenomenon of node anti--bunching.

Where is the attraction coming from? The answer is:
Interference of waves.
When we interfere infinitely many plane waves of
identical amplitudes and wave numbers, and allow all
propagation directions with equal weight,
the interference pattern is that
of the Bessel function $J_{0}$. This surprising feature
is just the physical interpretation of the Sommerfeld
integral representation

\begin{displaymath}
J_{0}(k\: r)=\frac{1}{2\pi}\int_{0}^{2\pi}d\theta \:
e^{ik\:r\: \sin \theta} 
\end{displaymath}
of the Bessel function.

The particle represented by the wave function containing
$J_{0}$ feels the quantum
anti-centrifugal force.
Each plane wave contributing to the Bessel interference
pattern does not feel any force.
The interference of all plane waves acts as an effective
force. Attraction from interference!

\subsection{Negative energy}

So far, we have focused on the case of
positive energies. An interesting selection of
solutions occurs when we make the transition from
positive to negative energies. Due to the
sign change of the energy and the quadratic dispersion
relation of the free particle, the wave number becomes
purely imaginary. Consequently, the ordinary Bessel
functions $J_{m}$ and $Y_{m}$ turn into the
modified Bessel functions $I_{m}$ and $K_{m}$.
However, based on physical arguments, no solutions
of negative energy exist for $m\ge 1$. Nevertheless,
for $m=0$ we have the two solutions $I_{0}$ and $K_{0}$.
Since the modified Bessel function $I_{0}$
increases exponentially for large distances,
whereas $K_{0}$ decreases, the boundary conditions imposed
by the quantum anti-centrifugal potential select
$K_{0}$ and thus the solution Eq.(\ref{eq1}).

The reduction from two equally contributing waves to
a single one as we cross the zero energy line is reminiscent of
the behaviour of the Airy function when we cross \cite{ber}
the Stokes line going from negative to positive
arguments. Indeed, for large negative values we can
approximate the Airy function by two counterpropagating
waves, whereas for large positive values we
only find a single decaying exponential.

\subsection{Higher dimensions}

This phenomenon of attraction
is unique to two dimensions \cite{nieto}. Indeed, for a free
particle of vanishing angular momentum
the $N$-dimensional, (hyper)spherical Schr\"{o}dinger
equation \cite{gur}

\begin{displaymath}
\left\{ \frac{d^2}{dr^2}+\frac{2M}{\hbar^2}
\left[ E-V_{0}^{(N)}
\right] \right\} u^{(N)}(r)=0
\end{displaymath}
for the radial variable $r \equiv \left(
x_{1}^{2}+\ldots +x_{N}^{2}\right)^{1/2}$
contains the quantum potential \cite{com}

\begin{displaymath}
V_{0}^{(N)}(r)\equiv \frac{\hbar^2}{2M}\frac{(N-1)(N-3)}{4r^2}.
\end{displaymath}
For $N=1$ and $N=3$ the quantum potential vanishes.
For higher dimensions $N\ge 3$ it
is repulsive. Only for $N=2$ this potential
becomes attractive. Therefore, the anti-centrifugal
force effect is a consequence of the dimensionality
of space.

\section{Bound state of a `free' particle}
\label{boun}

These considerations suggest that in two dimensions
there exists a bound state of a free particle
with the wave function given by Eq.(\ref{eq1}).
However, we emphasize that the wave number $k$ and
thus the energy $E$ are free parameters. There
is no length scale in the problem.
What fixes the energy of this bound state?
The logarithmic singularity
of $\Phi^{(2)}$ at the origin indicates that
there the wave function does not satisfy
the time independent
Schr\"{o}dinger equation. Indeed, the
wave function Eq.(\ref{eq1}) satisfies the equation
\cite{ree}

\begin{displaymath}
\left[ \Delta^{(2)} -k^2\right] \Phi^{(2)}(\vec{r})=
U_{0}\: \delta^{(2)}(\vec{r})
\end{displaymath}
with an additional delta function potential \cite{fer} of strength
$U_{0}$. A nonlinear relation between $k$ and $U_{0}$
determines \cite{wod} the eigen energy of the bound state.

Hence we are not really dealing with a free particle,
but with a particle in the presence of a delta function
potential. Notwithstanding
the problems \cite{wod,alb} associated with the definition of a delta
function potential in two and higher dimensions, it is well known
that under appropriate conditions such potentials entertain
bound states \cite{wod,alb}. Indeed, in one dimension the strength
$U_{0}$ of the potential has to be negative and
the corresponding probability distribution

\begin{eqnarray*}
W^{(1)}(x)\: dx & \equiv & \mid \Phi^{(1)}(x) \mid^2 dx
= \left( \sqrt{k}
e^{-k \mid x \mid}\right)^2 dx
\nonumber \\
& = & ke^{-2k\mid x\mid} \: dx
\end{eqnarray*}
displays a maximum at the location of the potential.

In three dimensions the parameter $U_{0}$ has to be positive
in order for the delta function potential to support
a bound state. As in one dimension, the probability
distribution

\begin{eqnarray*}
W^{(3)}(r)\: dr & \equiv &
\mid \Phi^{(3)}(r)
\mid^2 4\pi r^2 dr \nonumber \\
& = &
\left( \sqrt{\frac{k}{2\pi}}\frac{1}{r}e^{-k r}
\right)^2 4\pi r^2 dr
=2k\:e^{-2kr}\:dr
\end{eqnarray*}
is an exponential and exhibits a maximum at the origin.

The reason for this common feature
is quite intriguing.
In one dimension it is simply due to the fact that the
wave function $\Phi^{(1)}(x)$
has a maximum at $x=0$. In three dimensions the
situation is more subtle. Here the radial
wave function $\Phi^{(3)}(r)$ contains a
$1/r$-singularity, creating a $1/r^2$-singularity
in the probability density. However, the volume
element $4\pi r^2dr$ of a spherical shell in three
dimensions cancels the singularity in the probability
and only the exponential at the origin survives.

In two dimensions the situation is drastically different.
Independent of the sign of $U_{0}$ there always exists
a single bound state, with wave function $\Phi^{(2)}$, Eq.(\ref{eq1}).
Moreover, the area element $2\pi r\: dr$ of a ring prevails
over the logarithmic singularity contained in $K_{0}$.
This creates a node at the origin.
As a consequence, the maximum of the probability distribution
gets pushed away from the center of attraction.

In this sense, the intuitive picture of a repulsive
centrifugal force reappears: The maximum of the probability
is not at the origin, but in a ring surrounding it.
The quantum anti-centrifugal potential keeps the
packet together.

This behaviour is reminiscent
of the probability distribution of the electron in the
hydrogen atom, in a $s$ state. Here, the wave function is
an exponential and displays a maximum at the origin.
The volume element $4 \pi r^2 \: dr$ of a three-dimensional
spherical shell creates a
node at the origin and thus a maximum
at the Bohr radius. However, there is a fundamental difference to our
situation: The exponential decay of the wave function in the atom
is enforced by a classical potential, namely the
Coulomb potential. In contrast, for the free particle in two
dimensions it is the quantum anti-centrifugal potential
which demands the decay.

\section{Conclusions}
\label{conc}

There is an interesting connection between the
energy eigenstates of a free particle in two dimensions
and diffraction-free beams \cite{dur},
that is Bessel beams \cite{mug} in classical optics.
Here the ordinary Bessel function
$J_{0}$ describes the wave field with a purely real
wave number corresponding to positive energy.
However, the present effect
corresponds to negative energies and relies on purely imaginary
wave numbers giving rise to modified Bessel functions.
This is analogous to axicons used in classical optics.

This phenomenon of binding a particle with the help
of the quantum anti--centrifugal force
could have interesting applications in
the context of waveguides. Needless to say, all the
conclusions hold for electromagnetic fields
when we can ignore polarization. Here the maximum
of the intensity does not lie in the waveguide, that
is the delta function potential, but rather outside.

The newly emerging field of cold atoms offers interesting
possibilities for experimentally verifying the existence
of the quantum anti--centrifugal force. Here we
do not go into the details of such an experiment,
but only give an idea. The interaction
between two cold atoms is usually modelled by a delta
function. We can use this feature to create the delta--function
potential necessary for the wave function $\Phi^{(2)}$
defined in Eq. (\ref{eq1}) to be an eigenstate of the
self--adjoint extension of the kinetic energy operator.
The cylindrical symmetry we achieve by
using a dilute atomic beam guided by a laser beam.
In order not to affect the atom to be trapped, we have to
work with two different atomic elements. In the sense
of a Born--Oppenheimer approximation the atom feels a
time--averaged delta function potential.

We conclude by emphasizing that this phenomenon
of attraction in a free particle crucially depends
on the fact that we have restricted the space
to two dimensions. For positive energies the special
case of vanishing angular momentum selects the
origin as a special point of the two dimensional
plane. In the case of negative energies with a delta
function potential the origin becomes a singular
point, much in the spirit of the singularity
provided by the magnetic flux line in the Aharonov--Bohm
effect. These facts demonstrate that
in two dimensions a single point matters:
It changes the topology. In contrast, in three dimensions
a single point is less important.

\section{Acknowledgements}

We thank G. Alber, H. Carmichael, R. Chiao, J. P. Dahl, J. H. Eberly,
B.-G. Englert, W. G\"{o}tze, Th. H\"{a}nsch, W. Jakubetz, D. Meschede,
M. M. Nieto, G. Raithel, J. Ruostekoski, H. Walther, K. W\'{o}dkiewicz,
H. Woerdman and C. Zimmermann for many fruitful discussions.
Two of us (KRz and WPS) thank the University of Texas at Austin
for the hospitality where part of this work was done.
The work of W. P. S. is partially supported by DFG.

\pagebreak

\onecolumn
\widetext

\pagebreak

\end{document}